\documentstyle[12pt]{article}

\def\hybrid{\topmargin -20pt    \oddsidemargin 0pt
        \headheight 0pt \headsep 0pt
        \textwidth 6.25in       
        \textheight 9.5in       
        \marginparwidth .875in
        \parskip 5pt plus 1pt   \jot = 1.5ex}

\hybrid

\def\baselinestretch{1.2}

\catcode`\@=11

\def\marginnote#1{}
\newcount\hour
\newcount\minute
\newtoks\amorpm
\hour=\time\divide\hour by60
\minute=\time{\multiply\hour by60 \global\advance\minute by-\hour}
\edef\standardtime{{\ifnum\hour<12 \global\amorpm={am}%
        \else\global\amorpm={pm}\advance\hour by-12 \fi
        \ifnum\hour=0 \hour=12 \fi
        \number\hour:\ifnum\minute<10 0\fi\number\minute\the\amorpm}}
\edef\militarytime{\number\hour:\ifnum\minute<10 0\fi\number\minute}

\def\draftlabel#1{{\@bsphack\if@filesw {\let\thepage\relax
   \xdef\@gtempa{\write\@auxout{\string
      \newlabel{#1}{{\@currentlabel}{\thepage}}}}}\@gtempa
   \if@nobreak \ifvmode\nobreak\fi\fi\fi\@esphack}
        \gdef\@eqnlabel{#1}}
\def\@eqnlabel{}
\def\@vacuum{}
\def\draftmarginnote#1{\marginpar{\raggedright\scriptsize\tt#1}}

\def\draft{\oddsidemargin -.5truein
        \def\@oddfoot{\sl preliminary draft \hfil
        \rm\thepage\hfil\sl\today\quad\militarytime}
        \let\@evenfoot\@oddfoot \overfullrule 3pt
        \let\label=\draftlabel
        \let\marginnote=\draftmarginnote
   \def\@eqnnum{(\theequation)\rlap{\kern\marginparsep\tt\@eqnlabel}%
\global\let\@eqnlabel\@vacuum}  }

\def\preprint{\twocolumn\sloppy\flushbottom\parindent 2em
        \leftmargini 2em\leftmarginv .5em\leftmarginvi .5em
        \oddsidemargin -.5in    \evensidemargin -.5in
        \columnsep .4in \footheight 0pt
        \textwidth 10.in        \topmargin  -.4in
        \headheight 12pt \topskip .4in
        \textheight 6.9in \footskip 0pt
        \def\@oddhead{\thepage\hfil\addtocounter{page}{1}\thepage}
        \let\@evenhead\@oddhead \def\@oddfoot{} \def\@evenfoot{} }

\def\numberbysection{\@addtoreset{equation}{section}
        \def\theequation{\thesection.\arabic{equation}}}

\def\underline#1{\relax\ifmmode\@@underline#1\else
        $\@@underline{\hbox{#1}}$\relax\fi}

\def\titlepage{\@restonecolfalse\if@twocolumn\@restonecoltrue\onecolumn
     \else \newpage \fi \thispagestyle{empty}\c@page\z@
        \def\thefootnote{\fnsymbol{footnote}} }

\def\endtitlepage{\if@restonecol\twocolumn \else \newpage \fi
        \def\thefootnote{\arabic{footnote}}
        \setcounter{footnote}{0}}  

\catcode`@=12
\relax

\def\figcap{\section*{Figure Captions\markboth
        {FIGURECAPTIONS}{FIGURECAPTIONS}}\list
        {Figure \arabic{enumi}:\hfill}{\settowidth\labelwidth{Figure
999:}
        \leftmargin\labelwidth
        \advance\leftmargin\labelsep\usecounter{enumi}}}
 \relax
\def\tablecap{\section*{Table Captions\markboth
        {TABLECAPTIONS}{TABLECAPTIONS}}\list
        {Table \arabic{enumi}:\hfill}{\settowidth\labelwidth{Table
999:}
        \leftmargin\labelwidth
        \advance\leftmargin\labelsep\usecounter{enumi}}}
 \relax
\def\reflist{\section*{References\markboth
        {REFLIST}{REFLIST}}\list
        {[\arabic{enumi}]\hfill}{\settowidth\labelwidth{[999]}
        \leftmargin\labelwidth
        \advance\leftmargin\labelsep\usecounter{enumi}}}
 \relax
\makeatletter
\newcounter{pubctr}
\def\publist{\@ifnextchar[{\@publist}{\@@publist}}
\def\@publist[#1]{\list
        {[\arabic{pubctr}]\hfill}{\settowidth\labelwidth{[999]}
        \leftmargin\labelwidth
        \advance\leftmargin\labelsep
        \@nmbrlisttrue\def\@listctr{pubctr}
        \setcounter{pubctr}{#1}\addtocounter{pubctr}{-1}}}
\def\@@publist{\list
        {[\arabic{pubctr}]\hfill}{\settowidth\labelwidth{[999]}
        \leftmargin\labelwidth
        \advance\leftmargin\labelsep
        \@nmbrlisttrue\def\@listctr{pubctr}}}
 \relax
\makeatother
\newskip\humongous \humongous=0pt plus 1000pt minus 1000pt

\newif\ifdtup

\relax

\def\be{\begin{equation}}
\def\ee{\end{equation}}
\def\ba{\begin{eqnarray}}
\def\ea{\end{eqnarray}}

\def\no{\noindent}

\def\IR{\relax{\rm I\kern-.18em R}}

\def\IR{\relax{\rm I\kern-.18em R}}
\def\inv{^{\raise.15ex\hbox{${\scriptscriptstyle -}$}\kern-.05em 1}}

\begin{document}

\renewcommand{\theequation}{\arabic{equation}}

\newcommand{\beq}{\begin{equation}}
\newcommand{\eeq}[1]{\label{#1}\end{equation}}
\newcommand{\ber}{\begin{eqnarray}}
\newcommand{\eer}[1]{\label{#1}\end{eqnarray}}
\newcommand{\eqn}[1]{(\ref{#1})}
\begin{titlepage}
\begin{center}

\hfill hep--th/0511057\\
\vskip -.1 cm
\hfill CERN-PH-TH/2005-211\\
\vskip -.1 cm
\hfill October 2005\\

\vskip .4in

{\large \bf Geometric flows and (some of) their physical applications}\footnote{Based 
on an invited lecture at the Alexander von Humboldt foundation international 
conference on {\em Advances in Physics and 
Astrophysics of the 21st Century}, held from  
6 to 11 September 2005 in Varna, Bulgaria; to appear in a Supplement to the 
Bulgarian Journal of Physics}

\vskip 0.6in

{\bf Ioannis Bakas\footnote{Present (permanent) address: Department of Physics, 
University of Patras, 26500 Patras, Greece; e-mail: bakas@ajax.physics.upatras.gr}} 
\vskip 0.2in
{\em Theory Division, Department of Physics, CERN\\
CH-1211 Geneva 23, Switzerland\\
\footnotesize{\tt ioannis.bakas@cern.ch}}\\

\end{center}

\vskip .8in

\centerline{\bf Abstract}

\no
The geometric evolution equations provide new ways to address a variety of non-linear 
problems in Riemannian geometry, and, at the same time, they enjoy numerous 
physical applications, most notably within the renormalization group analysis of 
non-linear sigma models and in general relativity. They are divided into 
classes of intrinsic and extrinsic curvature flows.  
Here, we review the main aspects of intrinsic geometric flows driven by the  
Ricci curvature, in various forms, and explain the intimate relation between 
Ricci and Calabi flows on K\"ahler manifolds using the notion of super-evolution.
The integration of these flows on two-dimensional surfaces relies on the 
introduction of a novel class of infinite dimensional algebras with infinite 
growth. It is also explained in this context how Kac's $K_2$ simple Lie algebra 
can be used to construct metrics   
on $S^2$ with prescribed scalar curvature equal to the sum of any holomorphic 
function and its complex conjugate; applications of this special problem to 
general relativity and to a model of interfaces in statistical mechanics are also 
briefly discussed. 
\vfill
\end{titlepage}
\eject

\def\baselinestretch{1.2}
\baselineskip 16 pt
\noindent
The geometric evolution equations are parabolic systems that describe the 
deformation of metrics on Riemannian manifolds driven by their curvature 
in various forms. The continuous parameter $t$ of the evolution is 
typically called 
time. These equations arise in a variety of non-linear problems in physics 
and mathematics and led to ground breaking results in recent years. 
They are naturally divided into classes of intrinsic and extrinsic curvature 
flows. The first class refers to deformations driven by the intrinsic Ricci 
curvature tensor on a manifold, 
whereas the second class refers to deformations of 
submanifolds embedded in higher dimensional spaces that evolve 
by their extrinsic curvature. Here, we will only be concerned with the 
mathematical structure and physical applications of intrinsic curvature 
flows, and review, apart from the basic facts, some recent results 
on their integrability in two dimensions, 
\cite{bakas1, bakas2, bakas3, bakas4}, by focusing on the so called 
Ricci and Calabi flows. Selected applications to 
quantum field theory and general relativity will also be discussed 
among others. There is also a small number of new results and 
connections that are distributed in the text. 

The subject  of extrinsic curvature flows, which is equally interesting and 
much older, in certain respects, will not be included in this 
presentation. We only mention verbose, without formulae, the 
important role of the so called inverse mean curvature flow in 
general relativity. It was 
first introduced by Geroch, \cite{gero}, in an attempt to prove positive 
energy theorems and examine the 
issue of naked singularities within Penrose's program, \cite{penro}. 
Profound results were obtained recently in this direction by proving the 
Riemannian Penrose inequality on general grounds, \cite{huis}; see also 
reference \cite{bray} for a more popular account. Another example of 
extrinsic flow is the so called mean curvature flow, which was first 
introduced as model for the motion of grain boundaries in an 
annealing piece of metal, \cite{mulli}, and then put on firm mathematical 
basis in a monograph by Brakke, \cite{brakk}, and the work of many 
others that followed. Such deformations are 
typically associated to surface tension forces of embedded submanifolds 
and the ruling equation is a gradient flow for the corresponding area 
functional. It turns out that the mean curvature 
flow can be identified with the renormalization group 
equation of Dirichlet sigma models away from conformality, \cite{leigh}, 
and then applied to various models of boundary quantum field theory
of current interest, \cite{clip}. Further details about this interpretation, as well as 
generalizations with the Dirac-Born-Infeld action, will appear in a forthcoming 
publication, \cite{mine}. This ends the brief account of another interesting 
topic that will be excluded here. 

The Ricci flow constitutes the prime example of intrinsic curvature flows. It is 
a system of second order non-linear parabolic equations for the components of the 
metric on any Riemannian manifold $M$, which undergo continuous 
deformations driven by the Ricci curvature tensor according to  
\begin{equation}
{\partial \over \partial t} g_{\mu \nu} = - R_{\mu \nu} ~.  
\end{equation}
It defines an infinite dimensional dynamical system in superspace, which  
consists of all possible metrics 
on a given Riemannian manifold, and as such it is very hard to analyze in all 
generality. The short time existence of its solutions is guaranteed by the 
parabolic form of the dynamics, but there might be singularities that arise 
at finite time, as in the case of compact manifolds with strictly positive 
curvature metrics. In this case the space collapses and the flow becomes 
extinct. It is customary in the mathematics literature to work with the 
so called normalized Ricci flow using the mean scalar curvature $<R>$ on $M$,  
\begin{equation}
{\partial \over \partial t} g_{\mu \nu} = - R_{\mu \nu} + 
{<R> \over {\rm dim} M} ~ g_{\mu \nu} ~,   
\end{equation}
which, unlike the original form of the flow, preserves the total volume 
of space. Thus, it has a better chance to admit solutions that exist for 
sufficiently long time and converge asymptotically to constant curvature 
metrics, either positive or negative, depending on the topology of space.
For this reason it has been employed as major new tool to address a 
variety of non-linear problems in Riemannian geometry, and, most notably, 
for the uniformization of manifolds in two and three dimensions. Of course, 
the normalized Ricci flow is no other than the unnormalized flow, since 
the two are related to each other by suitable time reparametrization and 
rescaling of the metric by a function of time.  

The Ricci flow was first introduced in mathematics by Hamilton in the early  
eighties, \cite{ham}, although the subject has much longer history in physics, where it 
appeared in the renormalization group studies of non-linear sigma models 
in two dimensions with fields taking values in a Riemannian 
manifold $M$. The original computation was performed by Polyakov who only 
considered the class of $O(n)$ sigma models with target space metric that of 
the round sphere on $S^{n-1}$ for $n \geq 3$. The radius of the sphere is the 
only parameter of the theory, and, as such, it serves as the inverse of its 
coupling constant, i.e., $R \sim 1/g$. These two-dimensional models are not 
conformally invariant in the quantum regime, as it was found that the 
coupling constant runs, \cite{poly}, 
\begin{equation}
{1 \over {\tilde{g}}^2} = {1 \over g^2} + {n-2 \over 4\pi} {\rm log} 
{\tilde{\Lambda} \over \Lambda} 
\end{equation}
with respect to the world-sheet renormalization scale parameter 
($\sim \Lambda^{-1}$), to lowest 
order in perturbation theory. As a result, the beta function is negative 
for all $n \geq 3$ and the theory becomes asymptotically free 
in the ultra-violet regime, thus providing a toy model for the 
asymptotic freedom that governs the ultra-violet behavior of the gauge 
theory of strong interactions in four space-time 
dimensions, \cite{gross, polit}, and makes perturbation theory reliable 
for calculations.  In more 
geometrical terms, the result corresponds to a particular solution of the 
(unnormalized) Ricci flow equation, which describes a uniformly contracting 
round sphere $S^{n-1}$ with radius-square that diminishes linearly with 
$t = {\rm log} \Lambda^{-1}$; the flow starts from a very large (small curvature) 
sphere in the ultra-violet region $t \rightarrow - \infty$ and evolves 
continuously towards smaller sizes, until the sphere crashes to a point.          
Close to the singularity, however, the lowest order expression for the beta 
function is not reliable for higher order corrections and 
non-perturbative effects, in particular, may lead to a different physical 
interpretation, as they do.  

The computation of the perturbative beta function was subsequently extended by 
Friedan to encompass two-dimensional non-linear sigma models with generalized 
coupling given by the target space metric $g$ of arbitrary Riemannian manifolds. 
It follows, \cite{fried}, 
\begin{equation}
\Lambda^{-1} {\partial \over \partial \Lambda^{-1}} g_{\mu \nu} = 
- \beta (g_{\mu \nu}) = - R_{\mu \nu} + \cdots ~, 
\end{equation}
where the dots stand for higher order curvature corrections. The identification 
with the Ricci flow follows by setting $t$ equal to the renormalization group 
time ${\rm log} \Lambda^{-1}$. Here, we will not be concerned with the form of 
higher order corrections and instanton effects to the beta function; the Ricci flow 
will be taken at face value, as in the mathematics literature. 
This calculation led to the development of the sigma model approach 
to string theory, since the requirement of conformal invariance for the world-sheet 
quantum field theory gives rise to vacuum Einstein equation in target space, 
$R_{\mu \nu} =0$, plus higher order corrections if they are included. Further 
generalizations were also considered by appending other massless modes of the string, 
such as a dilaton and an anti-symmetric 
tensor field (torsion), to the action of the world-sheet sigma model. In this case
one obtains a generalized system of beta function equations whose fixed points 
provide consistent backgrounds for string propagation. We only note here that 
the effect of the dilaton manifests as reparametrization along the Ricci flow, 
according to the more general equation 
\begin{equation}
{\partial \over \partial t} g_{\mu \nu} = - R_{\mu \nu} + \nabla_{\mu} \xi_{\nu} 
+ \nabla_{\nu} \xi_{\mu} ~, 
\end{equation}
when $\xi$ is a gradient vector field. Finally, the renormalization group 
trajectories, away from fixed points, have been used to provide an off-shell 
formulation of the problem of tachyon condensation in closed string theory 
that accounts for the decay of an unstable vacuum to another more stable 
vacuum.    

The presence of torsion 
affects the formation of singularities along the renormalization group flow and 
deserves systematic study in the future. Here, we will not be concerned with it 
because most of our results will be confined to models with two-dimensional target 
spaces, where there can be no perturbative contributions to the beta functions due 
to torsion; any contributions of this kind appear through the corresponding field 
strength, which is a 3-form, that vanishes identically in two dimensions. 
However, one may also add instanton contributions to the beta function that capture 
the effect of topological torsion in two-dimensional target space, as in the 
case of the $O(3)$ sigma model with a $\theta$-term. It is well known by now that 
at $\theta = \pi$ the model becomes massless and its infra-red limit is governed by 
the $SU(2)_1$ Wess-Zumino-Witten model, i.e., a free boson on a circle with 
its radius held fixed at the enhanced $SU(2)$ symmetry point, \cite{affl}. 
This result also finds important applications to 
the explanation of certain aspects of the quantum Hall effect (see, for instance, 
\cite{pruis}, and references therein).  It may serve as starting point for 
working mathematicians to explore the resolution of singularities of the Ricci 
flow by instantons and study their implications to geometry. We intend to return 
to these issues elsewhere, since they have been omitted from our work so far.    
   
There is already vast literature in mathematics concerning the qualitative 
behavior of solutions and the formation of singularities along the 
(normalized) Ricci flow in low dimensions (see, for instance, 
\cite{ricci1, ricci2, ricci3}, 
and references therein). These studies play important role 
in three dimensions, in particular, and have been advanced towards a proof of 
Thurston's geometrization conjecture by the recent work of Perelman, \cite{perel}. 
It is a subject of great activity in mathematics, since many steps of the proposed 
proof are being analyzed and cross checked by the experts; see, for 
instance, \cite{popular2}, for a more popular account. Apart from 
these general considerations there are also some simple solutions of the Ricci 
flow in low dimensions that were obtained by consistent truncation to 
mini-superspace sectors and depend on a small number of moduli. 
In two dimensions, they include 
the sausage model as an axially symmetric deformation of the $O(3)$ sigma 
model, \cite{sausa}, and the decay of a cone $C/Z_n$ to the plane, 
\cite{gulp}. The latter solution, which generalizes 
the fundamental (Gaussian) solution of the heat flow equation, exhibits a 
delta function singularity in the curvature at some initial time, which dissipates 
completely in the infra-red region, as $t \rightarrow +\infty$. It also 
serves as a model to study localized tachyon condensation in closed sting theory. 
In three dimensions, there is the $O(4)$ sigma model generalization of the 
sausage deformation, \cite{fatee}, as well as various examples of 
mini-superspace deformations within the Bianchi classification of allowed 
isometries, \cite{jim1}. There are also numerical studies 
of different trajectories that depend on specific initial data, 
which provide preliminary evidence for critical behavior 
along the Ricci flow, \cite{jim2}, as in the problem of gravitational collapse. 
In fact, this analogy will become stronger if one finds an off-shell 
description of the gravitational collapse in four space-time dimensions in 
terms of the Ricci flow on three-dimensional Riemannian manifolds.    

Next, we turn the discussion to the Calabi flow, which, unlike the Ricci flow, 
is only defined on K\"ahler manifolds with 
complex coordinates $(z^a, {\bar{z}}^a)$. It has the following form, \cite{cala}, 
\begin{equation}
{\partial \over \partial t} g_{a \bar{b}} = {\partial^2 R \over \partial z^a 
\partial {\bar{z}}^b} ~,  
\end{equation}
using the scalar Ricci curvature, which in complex notation is  
\begin{equation}
R = g^{a \bar{b}} R_{a \bar{b}} = - g^{a \bar{b}}  
\partial_a {\bar{\partial}}_b {\rm log} ({\rm det} g) ~.  
\end{equation}
The Calabi flow is a fourth order equation for the 
components of the metric, and, thus, more 
complicated to study in all generality as compared to the second order Ricci flow.  
In fact, although several simple solutions are explicitly known for the Ricci 
flow, there are hardly any examples available in the literature for trajectories  
of the Calabi flow. In other words, it appears to be difficult to device  
mini-superspace models of the Calabi flow for which all relevant deformations 
can be consistently truncated to a finite number of moduli.    
Both  K\"ahler-Ricci and Calabi flows induce deformations of the metric in a 
fixed cohomology class, but the total volume of space is preserved in the latter  
case, unlike the unnormalized Ricci flow. Critical points of the Calabi flow are 
called extremal metrics. They only exit under certain technical conditions 
and often describe constant curvature metrics on a given K\"ahler manifold; 
see, for instance, \cite{tian}, for reviews of the subject. The restriction of 
geometric evolution equations to K\"ahlerian spaces is a great relief for  
the system of induced deformations, since there is only one function, 
namely the K\"ahler potential $K$, that determines the form of the metric as 
$g_{a \bar{b}} (z, \bar{z}) = \partial_a {\bar{\partial}}_b K(z, \bar{z})$. 
Yet, there are several applications to geometry that include new proofs of 
uniformization theorems and the investigation of some open conjectures for 
the curvature of K\"ahler manifolds.    

In the following, we restrict attention to geometric evolutions on two-dimensional 
surfaces and write the metric in a system of conformally flat (K\"ahler) 
coordinates $(z, \bar{z})$,  
\begin{equation}
ds_{\rm t}^2 = 2 e^{\Phi (z, \bar{z}; t)} dz d\bar{z} ~. 
\end{equation}
Then, the two different flows assume the simple form  
\begin{eqnarray}
{\rm Ricci} ~ {\rm flow} & : & ~~~~~~~ {\partial \Phi \over \partial t} = 
\Delta \Phi ~, \\
{\rm Calabi} ~ {\rm flow} & : & ~~~~~~~ {\partial \Phi \over \partial t} = 
- \Delta \Delta \Phi ~, 
\end{eqnarray}
where $\Delta$ is the Laplace-Beltrami operator on the surface, 
\begin{equation}
\Delta = e^{-\Phi} \partial \bar{\partial} ~. 
\end{equation}
These are non-linear differential equations of second and fourth order, respectively,  
that resemble the heat flow equation on the surface due to the parabolic 
character of 
their dependence on the time variable $t$. As such, they exhibit dissipative 
behavior in time, but, as will be seen later, 
they turn out to be integrable in space with the aid of 
appropriately chosen infinite dimensional Lie algebras that incorporate the 
deformation variable $t$ into their defining relations.   

It should be noted at this point, as side remark, that the two-dimensional 
geometric flows arise as limiting cases of more general diffusion equations on 
surfaces. First, let us consider the porous medium equation 
\begin{equation}
{\partial u \over \partial t} = \partial \bar{\partial} u^m ~,      
\end{equation}
where $m$ is an arbitrary (possibly fractional) exponent. The limit 
$m \rightarrow 0$ is well defined provided that $t$ is also rescaled by $m$ so that 
$mt$ remains finite. Then, in the new time variable, the evolution 
becomes (see, for instance, \cite{wu}) 
\begin{equation}
{\partial u \over \partial t} = \lim_{m \rightarrow 0} {1 \over m} 
\partial \bar{\partial} u^m = \partial \bar{\partial} {\rm log} u ~,      
\end{equation}
which is identical to the Ricci flow provided that $u = e^{\Phi}$. Likewise, 
there is a fourth order diffusion equation (with no name, to the best of our 
knowledge) of the form 
\begin{equation}
{\partial u \over \partial t} = - \partial \bar{\partial} (u^{-1} \partial 
\bar{\partial} u^m ) ~,      
\end{equation}
whose $m \rightarrow 0$ limit yields in a similar fashion the equation 
\begin{equation}
{\partial u \over \partial t} = - \partial \bar{\partial} (u^{-1} \partial 
\bar{\partial} {\rm log} u ) ~,      
\end{equation}
i.e., the two-dimensional Calabi flow for the variable $u=e^{\Phi}$. Some 
properties of the geometric evolution equations are inherited from these 
more general diffusion processes, although it is not clear to us whether 
the group theoretical methods used later for their integration also extend 
to the equations with arbitrary values of the exponent $m$.  

A physical context for the Calabi flow on two-dimensional surfaces with spherical 
topology is provided by the class of Robinson-Trautman metrics in 
general relativity. These are radiative metrics in four space-time dimensions of the 
form, \cite{robi, exact}  
\begin{equation}
ds^2 = 2r^2 e^{\Phi(z, \bar{z}, t)} dz d\bar{z} -2dt dr - 
H(z, \bar{z}, r, t) dt^2  
\end{equation}
and they represent spherical gravitational waves in vacuum. The time component of the 
metric takes the form 
\begin{equation}
H = r {\partial \Phi \over \partial t} - \Delta \Phi - {2m \over r} ~,  
\end{equation}
where $m$ is a mass parameter that can be taken to be constant. In this context, 
Einstein's equations amount to a single differential equation for the conformal 
factor of closed surfaces with constant $r$ and $t$, which reads  
\begin{equation}
\Delta \Delta \Phi + 3m {\partial \Phi \over \partial t} = 0 ~.  
\end{equation}
When $m > 0$, it is identical to the Calabi flow with deformation parameter $t$ 
given by the retarded time, \cite{tod}. Without loss of generality, one may set $3m=1$ in 
appropriate units. In this case, the most general solution describes type II 
space-times that evolve towards the Schwartzschild metric (see 
\cite{piotr} and references therein for earlier work on the 
subject). Thus, it becomes very  
important to explore the exact solvability of this particular dynamical sector
of Einstein's equation in terms of the two-dimensional Calabi flow. Higher 
dimensional generalizations of this flow have not yet appeared in physical 
problems to the best of our knowledge.   

The time evolution has the tendency to dissipate away any curvature perturbations 
of the canonical metric on the two-dimensional surface. 
In fact, after sufficiently long time, the 
metrics deform towards the constant curvature metric on the surface, which for the 
case of $S^2$ is $ds^2 = R_0^2 (d\theta^2 + {\rm sin}^2 \theta d\phi^2)$ in a system 
of spherical coordinates $(\theta, \phi)$. Axially symmetric deformations of a round  
sphere with radius $R_0$ are conveniently described in the form   
\begin{equation}
ds_{\rm t}^2 = R_0^2 [1+ \epsilon_l (t) P_l ({\rm cos} \theta)] 
\left(d\theta^2 + {\rm sin}^2 \theta d \phi^2 \right) ,  
\end{equation}
using Legendre polynomials $P_l({\rm cos}\theta)$ with $l \geq 2$. These deformations 
preserve the volume of the space. Then, to linear order in the perturbation parameters 
$\epsilon_l(t)$, the Calabi flow yields the following evolution  
\begin{equation}
\epsilon_l (t) = \epsilon_l(0) {\rm exp} \left(-{t \over 4R_0^4} l(l^2 -1) 
(l+2) \right) 
\end{equation}
that approximates well the asymptotic behavior of the full non-linear evolution as 
$t \rightarrow +\infty$. The result implies that the perturbations are damped 
exponentially fast for all $l \geq 2$ and the configuration settles down to that of a 
round metric on $S^2$. The damping is faster for higher values of $l$. In the 
context of Robinson-Trautman metrics, these perturbations correspond to 
linearized multi-pole gravitational radiation, \cite{ted}, 
as the geometry tends asymptotically to the Schwartzschild metric in 
Eddington-Finkelstein coordinates.   

Direct comparison with the dissipative properties of the Ricci flow on $S^2$ requires 
making use of the normalized (rather than the unnormalized) form of the evolution   
\begin{equation}
{\partial \Phi \over \partial t} = \Delta \Phi + {1 \over R_0^2} 
\end{equation}
so that the volume of the space is preserved. Then, the ansatz of infinitesimal 
axially symmetric deformations can be consistently implemented, and, to linear 
order in $\epsilon_l(t)$, it follows that  
\begin{equation}
\epsilon_l (t) = \epsilon_l(0) {\rm exp} \left(-{t \over 2R_0^2} (l-1) 
(l+2) \right) .  
\end{equation}
The perturbations are also damped exponentially fast, but at slower pace as compared  
to the fall off rates of the Calabi flow. Their specific dependence on $l$ reflects the 
difference in the order of the corresponding differential equations. Also, in either case, 
it is observed that the decay of perturbations depends on the radius $R_0$ and it is  
bigger for smaller spheres.   

There is a formal relation between Ricci and Calabi flows on K\"ahler manifolds, which is 
seen by taking the time derivative of the Ricci flow in complex coordinates, \cite{bakas4}, 
\begin{equation}
{\partial^2 \over \partial t^2} g_{a \bar{b}} = - {\partial \over \partial t} R_{a \bar{b}} 
= \partial_a \bar{\partial}_b {\partial \over \partial t} ({\rm log} ~ {\rm det} g) 
= - \partial_a \bar{\partial}_b R ~.  
\end{equation}
Thus, if the second derivative of the K\"ahler metric with respect to the Ricci time 
$t_{\rm R}$ is 
identified with minus its first derivative with respect to the Calabi 
time $t_{\rm C}$, i.e., 
\begin{equation}
{\partial^2 \over \partial t_{\rm R}^2} = - {\partial \over \partial t_{\rm C}} ~, 
\end{equation}
the two flows will be formally the same. This observation relates two parabolic 
equations of second and fourth order and it is reminiscent of the way to extract the 
square root of the Schr\"odinger equation in supersymmetric quantum mechanics. It 
will be particularly useful for the parallel treatment of the two flows on two 
dimensional surfaces.   

The integration of Ricci and Calabi flows beyond the linearized approximation relies 
on the use of appropriately chosen infinite dimensional algebras that enable to cast the 
equations into zero curvature form in two dimensions. The treatment of the two flows 
will be similar, as they both admit Toda field theoretic interpretation with gauge 
connections taking values in the class of (super)-continual Lie algebras for 
specific choices of the Cartan operator.   
In preparation of this, it is useful to implement the formal relation between the Ricci and 
Calabi flows on K\"ahler manifolds using the notion of super-evolution. In particular, 
let us introduce an anti-commuting variable $\theta$, with $\theta^2 = 0$, 
as the supersymmetric 
partner of the deformation variable $t$, and write down the super-evolution 
Ricci type equation, \cite{bakas4}, 
\begin{equation}
{\cal D}_T {\rm exp} {\cal F} = \partial \bar{\partial} {\cal F} ~, 
\end{equation}
where 
\begin{equation}
{\cal F}(z, \bar{z}; T) = \Phi(z, \bar{z}; t) + \theta \Psi (z, \bar{z}; t) 
\end{equation}
is a superfunction in $R^{1|1}$ superspace with coordinates $T=(t, \theta)$ and ${\cal D}_T$ 
is the associated super-derivative operator 
\begin{equation}
{\cal D}_T = {\partial \over \partial \theta} - \theta {\partial \over \partial t} 
\end{equation}
that satisfies the relation ${\cal D}_T^2 = - \partial / \partial t$. The components 
$\Phi$ and $\Psi$ are ordinary (bosonic) functions of $t$ that are also taken to depend on 
the complex coordinates $(z, \bar{z})$ on the surface. Then, in terms of components, the 
super-evolution equation for ${\cal F}$ reads 
\begin{equation}
e^{\Phi} \Psi = \partial \bar{\partial} \Phi ~, ~~~~~ {\partial e^{\Phi} \over \partial t} 
= - \partial \bar{\partial} \Psi ~, 
\end{equation}
which leads to the two-dimensional Calabi flow for $\Phi$ by eliminating the field $\Psi$.   
Thus, taking the square root of the time derivative operator, allows to connect the two 
distinct classes of geometric deformations of second and fourth order, respectively, via 
super-evolution, as was anticipated before on general grounds.   

The algebraic framework that is appropriate to use for integrating the Ricci and 
Calabi flows in two dimensions is provided by the class of 
(super)-continual Lie algebras with basic system of commutation relations, \cite{misha},  
\begin{eqnarray}
& & [H(\varphi) , ~ X^{\pm}(\psi)] = \pm X^{\pm} ((K\varphi) \psi) ~, \\
& & [X^+(\varphi) , ~ X^-(\psi)] = H (S(\varphi \psi)) ~, \\
& & [H(\varphi) , ~ H(\psi)] = 0 ~.  
\end{eqnarray}
The elements $H$ and $X^{\pm}$ generate the local part of the algebra, whereas all other
generators can be obtained, in principle,  
by taking successive commutators of these basic elements. 
The smearing functions can be either ordinary functions of the continuous variable $t$,  
in which case the corresponding Lie algebra is called continual, or superfunctions 
of the super-variable $T$, in which case we will refer to it as super-continual Lie 
algebra, \cite{bakas4}.
In either case, the algebras are bosonic, infinite dimensional, and they are solely  
characterized by the choice of operators $K$ and $S$ that act linearly on the space 
of smearing (super)functions. For the case of invertible operators, it is possible to 
make the canonical choice $S=1$ by redefining $K$ as $\tilde{K} = KS$. This choice 
will be adopted in the following, unless it is explicitly stated otherwise, and 
$\tilde{K}$ will be called Cartan operator; the tilde will also be dropped for simplicity.

Alternatively, one may consider a system of basic generators $H(t)$ and $X^{\pm}(t)$ 
which are related to the smeared ones as 
\begin{equation}
A(\varphi) = \int A(t) \varphi (t) dt  
\end{equation}
when $A(t)$ is smeared with any smooth function $\varphi (t)$ with compact 
support. In this formulation, the continual Lie algebra is characterized by the choice 
of Cartan kernel $K(t, t^{\prime})$, which is in general a distribution. For  
super-continual Lie algebras, the generators can be alternatively thought to depend on 
$T$, so that $A(T) = A_0(t) + \theta A_1(t)$, and their smearing is simply defined  
by integration in $R^{1|1}$ superspace against any suitably chosen superfunction 
${\cal F}(T) = \varphi_0(t) + \theta \varphi_1 (t)$, as  
\begin{equation}
A({\cal F}) = \int A(T) {\cal F} (T) dT = A_0 (\varphi_1) + A_1 (\varphi_0) ~,    
\end{equation}
using the relations $\int d \theta = 0$ and $\int \theta d \theta = 1$. 
Thus, it is possible to work out the form of the basic commutation relations 
of the Lie algebra in terms of bosonic components $H_i$ and $X^{\pm}_i$ with 
$i=0, 1$ in smeared or unsmeared form, if desired.   

For any given choice of Cartan operator $K$ there is an associated Toda equation 
for the field $\Phi (z, \bar{z}; t)$  
\begin{equation}
\partial \bar{\partial} \Phi = K(e^{\Phi}) 
\end{equation}
that admits zero curvature representation in two dimensions, 
\begin{equation}
[\partial + A_{+}(z, \bar{z}) , ~ \bar{\partial} + A_- (z, \bar{z})] = 0 ~,  
\end{equation}
using gauge connections $A_{\pm}$ with values in the corresponding continual Lie algebra.
Indeed, the particular choice 
\begin{equation}
A_+ = H(g) + X^+(1) ~, ~~~~~~ A_- = X^-(e^{\Phi})  
\end{equation}
amounts to the following system of equations 
\begin{equation}
\bar{\partial} g = e^{\Phi} ~, ~~~~~~ \partial \Phi = K(g) ~, 
\end{equation}
which yield the Toda field equation above after eliminating $g$. In the case of 
super-continual Lie algebras the framework is the same provided that both $\Phi$ and 
$g$ are superfunctions of $T$ rather than ordinary functions of $t$.   
 
It is becoming clear now that the two-dimensional Ricci flow admits the following Toda field 
theory interpretation 
\begin{equation}
{\partial e^{\Phi} \over \partial t} = \partial \bar{\partial} \Phi ~ : ~~~~~ 
K = {\partial \over \partial t} ~,  
\end{equation}
using a continual Lie algebra with Cartan operator $K = \partial / \partial t$, 
\cite{bakas1, bakas2}, whereas 
for the Calabi flow a similar interpretation follows easily from its equivalent description 
as super-evolution equation  
\begin{equation}
{\cal D}_T {\rm exp} {\cal F} = \partial \bar{\partial} {\cal F}  ~ : ~~~~~ 
K = {\cal D}_T = {\partial \over \partial \theta} - \theta {\partial \over \partial t} ~,  
\end{equation}
using a super-continual Lie algebra with odd Cartan operator $K= {\cal D}_T$ equal to the 
square root of $-\partial / \partial t$, \cite{bakas4}. 
In either case, the main idea of this formulation  
is to treat the coordinates $(z, \bar{z})$ and $t$ in uneven way and incorporate 
the deformation variable into the defining relations of an infinite dimensional 
(super)-continual Lie algebra. Then, the resulting Toda field theory interpretation of 
the flows allows for their zero curvature formulation in two dimensions, and,   
eventually, for their (formal) integration by group theoretical methods, as outlined below.    

The general solution of Toda field equations with Cartan operator $K$ is obtained 
by straightforward generalization of the group theoretical construction that is available 
for ordinary Toda systems, \cite{misha}. 
Let us introduce a highest weight state $|t>$ that depends on the 
continuous variable $t$ and satisfies the conditions    
\begin{equation}
X^+(t^{\prime}) |t> = 0 ~, ~~~~~ <t| X^-(t^{\prime}) = 0 ~, ~~~~~ 
H(t^{\prime}) |t> = \delta (t-t^{\prime}) |t> 
\end{equation}
and $<t|t> = 1$. The existence of such state is not guaranteed for continual Lie algebras,  
but, nevertheless, it will be formally introduced for all $K$. We also consider a 
one-parameter family (parametrized by $t$) of two-dimensional free fields given by the 
sum of an arbitrary holomorphic function and its complex conjugate,   
\begin{equation}
\Phi_0 (z, \bar{z}; t) = f(z; t) + \bar{f} (\bar{z}; t) ~,   
\end{equation}
and define the path-ordered exponentials 
\begin{eqnarray}
M_+ (z) & = & {\cal P} {\rm exp} \left( \int^z dz^{\prime} \int dt^{\prime} 
e^{f(z^{\prime}; t^{\prime})} X^+ (t^{\prime}) \right) , \nonumber\\ 
M_- (\bar{z}) & = & {\cal P} {\rm exp} \left( \int^{\bar{z}} d{\bar{z}}^{\prime} 
\int dt^{\prime} 
e^{\bar{f}({\bar{z}}^{\prime}; t^{\prime})} X^- (t^{\prime}) \right)  
\end{eqnarray}
obtained by suitable smearing of the generators $X^{\pm}$ with the holomorphic and 
anti-holomorphic components of $\Phi_0$, respectively. Then, the Toda field configurations 
admit, in general, the following free field representation  
\begin{equation}
\Phi (z, \bar{z}; t) = \Phi_0 (z, \bar{z}; t) - K \left({\rm log} <t| M_+^{-1}(z) 
M_- (\bar{z}) |t> \right) ~.  
\end{equation}

The expression above gives rise to an infinite power series by expanding 
the path-ordered exponentials $M_{\pm}$ and computing the expectation value  
\begin{eqnarray}
& & <t|M_+^{-1} M_- |t> = 1 + \sum_{n=1}^{\infty} (-1)^n \int^z dz_1 \cdots 
\int^{z_{n-1}} dz_n \int^{\bar{z}} d{\bar{z}}_1 \cdots \int^{{\bar{z}}_{n-1}} 
d{\bar{z}}_n \times \nonumber\\ 
& & ~~~~~~~~ \times \int \prod_{i=1}^n dt_i \int \prod_{i=1}^n dt_i^{\prime} 
{\rm exp} f(z_i; t_i) {\rm exp} \bar{f} ({\bar{z}}_i; t_i^{\prime}) 
D_t^{\{t_1, t_2, \cdots , t_n; t_n^{\prime}, \cdots , t_2^{\prime}, t_1^{\prime}\}}  
\end{eqnarray}
with $z \geq z_1 \geq \cdots \geq z_n$ and likewise for the $\bar{z}$'s. All  
information about the algebra is fully encoded into the elements 
\begin{equation}
D_t^{\{t_1, t_2, \cdots , t_n; t_1^{\prime}, t_2^{\prime}, \cdots , t_n^{\prime}\}} 
= <t| X^{+}(t_1) X^+ (t_2) \cdots X^+(t_n) X^-(t_n^{\prime}) \cdots X^-(t_2^{\prime}) 
X^-(t_1^{\prime}) |t> 
\end{equation}
that can be computed recursively by shifting $X^+$'s to the right and $X^-$'s to the 
left. Provided that the infinite series converges, the result is interpreted as a 
formal expansion of the non-linear configuration $\Phi$ around the free field $\Phi_0$
in powers of ${\rm exp} \Phi_0$.  

According to this framework, the general solution of the two-dimensional Ricci flow 
follows by specialization to the continual Lie algebra with 
Cartan operator $K = \partial / \partial t$. It is an example of a novel infinite 
dimensional algebra with anti-symmetric Cartan kernel that exhibits infinite growth 
beyond its local part, \cite{misha}.
Particular details of the calculation and some simple examples can be found in 
the literature, \cite{bakas1}, where the validity of the group 
theoretical integration has also been 
tested using exact mini-superspace models of axially symmetric deformations of constant 
curvature metrics. For example, axially symmetric configurations with free fields of the 
form $\Phi_0 = c \cdot (z+ \bar{z}) + d(t)$ admit the expansion in powers of 
${\rm exp} \Phi_0$, 
\begin{equation}
\Phi(z, \bar{z}; t) = \Phi_0 + {1 \over (1! ~c)^2} \partial_t e^{\Phi_0} 
+ {1 \over (2! ~c^2)^2} \partial_t \left(e^{\Phi_0} 
\partial_t e^{\Phi_0}\right) + {\cal O}(e^{3\Phi_0}) ~.  
\end{equation}
 
Likewise, the general solution of the two-dimensional Calabi 
flow (in superfield form) follows by specialization to the super-continual Lie algebra  
with odd Cartan operator $K={\cal D}_T$, as 
\begin{equation}
{\cal F} (z, \bar{z}; T) = {\cal F}_0 (z, \bar{z}; T) - K \left({\rm log} <T| M_+^{-1}(z) 
M_- (\bar{z}) |T> \right) .  
\end{equation}
The corresponding normalized highest weight state $|T>=  
|t>_0 + \theta |t>_1$ is introduced here using the smeared form of the defining     
relations,   
\begin{equation}
X^+(\varphi) |\psi> = 0 ~, ~~~~~ <\psi| X^-(\varphi) = 0 ~, ~~~~~ 
H(\varphi) |\psi> = |\varphi \psi>  ,  
\end{equation}
for all superfunctions $\varphi$ and $\psi$, and 
$|\psi> = |\psi_1>_0 + |\psi_0>_1$. 
Also, the free superfield is decomposed into sum of an arbitrary holomorphic superfunction 
and its complex conjugate, which are 
subsequently used to smear $X^{\pm}(T)$ in the corresponding 
expressions of the path-ordered exponential $M_{\pm}$. It turns out that axially 
symmetric deformations which correspond to free field configurations ${\cal F}_0 = 
\Phi_0 + \theta \Psi_0$ with $\Phi_0 = c \cdot (z+\bar{z}) + d(t)$ and 
$\Psi_0 = a \cdot (z+\bar{z}) + b(t)$, admit the following power series expansion 
\begin{equation}
\Phi = \Phi_0 + {1 \over (1! ~c)^2} \left(\Psi_0 - {2a \over c}\right) e^{\Phi_0} 
-{1 \over (2! ~c^2)^2} \left(e^{\Phi_0} \partial_t e^{\Phi_0} - \left(\Psi_0^2 
-{4a \over c} \Psi_0 + {7a^2 \over 2c^2} \right) e^{2\Phi_0} \right)  
\end{equation}      
plus higher terms of order ${\cal O}(e^{3\Phi_0})$. We refer the reader to the literature for 
further technical details, \cite{bakas4}.

We are turning now the discussion to a special topic related to static configurations 
of the two-dimensional Calabi flow, namely $\partial \bar{\partial} R = 0$, which amount 
to the following fourth order equation for the conformal factor, 
\begin{equation}
\Delta \Delta \Phi = 0 ~.  
\end{equation}
Constant curvature metrics arise as special solutions and they correspond to the 
Liouville equation, $\Delta \Phi = {\em const.}$, which admits zero curvature formulation 
in terms of $SL(2)$-valued gauge connections; its general solution is described using an 
arbitrary holomorphic function (and its complex conjugate) by specializing the 
integration scheme for Toda systems to the algebra with Cartan operator $K=1$.  
Constant curvature metrics also describe the static configurations of the 
normalized Ricci flow in two dimensions. 
More generally, however, solutions of the fourth order equation correspond to 
conformally flat metrics with prescribed scalar curvature equal to the sum of an 
arbitrary holomorphic function and its complex conjugate, i.e., 
\begin{equation}
\Delta \Phi = \psi(z) + \bar{\psi} (\bar{z}) ~.  
\end{equation}
A particularly simple solution, in fact, the only known example of this kind with 
$\Delta \Phi = z+ \bar{z}$, is 
\begin{equation}
e^{\Phi} = {3 \over (z+ \bar{z})^3} ~. 
\end{equation}
A well known physical framework for this equation is provided by the class of type III 
Robinson-Trautman metrics in general relativity, i.e., metrics with mass 
parameter $m=0$, \cite{robi, exact}. 

An alternative description of the equation $\Delta \Delta \Phi = 0$ is obtained in 
the form, \cite{bakas4},  
\begin{equation}
\partial \bar{\partial} {\cal F} = {\partial \over \partial \theta} e^{{\cal F}} 
\end{equation}
using a superfunction ${\cal F} (z, \bar{z}; \theta) = \Phi (z, \bar{z}) + 
\theta \Psi (z, \bar{z})$ with components related to each other as 
$\Psi = \Delta \Phi$. This formulation is advantageous for the formal integration 
of the equation by group theoretical methods, as in Toda systems. It turns out that 
the appropriate choice is the super-continual Lie algebra with $K=1$ and 
$S= \partial / \partial \theta$ in the nomenclature introduced earlier. 
There is no canonical choice of Cartan operator in this special case, since 
$S^2 = 0$; also the dependence of the algebra generators upon $t$ is superfluous and 
will be dropped out completely, thus leaving only their dependence on the Grassmann 
variable $\theta$.    
For later reference, it is useful to present the form of the commutation 
relations among the basic generators in terms of components,   
\begin{eqnarray}
& & [H_1 , ~ X_0^{\pm}] = \pm X_0^{\pm}  ~, ~~~~~ 
[X_1^{\pm}  , ~ X_0^{\mp} ] = \pm H_1  ~, \nonumber\\ 
& & [H_0 , ~ X_1^{\pm}] = \pm X_0^{\pm} ~, ~~~~~ 
[H_1 , ~ X_1^{\pm}] = \pm X_1^{\pm} ~,  
\end{eqnarray}
whereas the rest are trivial, 
\begin{eqnarray}
& & [X_0^+ , ~ X_0^-] = 0 ~, ~~~~~ 
[X_1^+ , ~ X_1^- ] = 0 ~, \nonumber\\ 
& & [H_0 , ~ X_0^{\pm}] = 0 ~, ~~~~~ 
[H_i , ~ H_j ] = 0 ~.  
\end{eqnarray}
It can be verified that the zero curvature condition for the gauge connections 
\begin{equation}
A_+ = f H_0 + g H_1 + X_1^+ ~, ~~~~~~  
A_- = \Psi e^{\Phi} X_0^-  + e^{\Phi} X_1^-  
\end{equation}
amounts to $\Delta \Delta \Phi = 0$ by eliminating all other functions but $\Phi$ 
from the resulting system of equations.  

The generators $H_1$ and $X_i^{\pm}$ form the local part of Kac's $K_2$ simple 
Lie algebra, as can be seen by setting  
\begin{eqnarray}
& & e_0 = {1 \over \sqrt{2}}(X_0^+ + X_1^+) ~, ~~~~~~ 
e_1 = {1 \over \sqrt{2}}(X_0^+ - X_1^+) ~, \nonumber\\ 
& & f_0 = {1 \over \sqrt{2}}(X_0^- + X_1^-) ~, ~~~~~~ 
f_1 = {1 \over \sqrt{2}}(-X_0^- + X_1^-) ~.   
\end{eqnarray}
Indeed, by also denoting $H_1 = h$, it follows that  
\begin{equation}
[e_i , ~ f_j] = \delta_{ij} h ~, ~~~~~ [h , ~ e_i] = e_i ~, ~~~~~ 
[h, ~ f_i] = -f_i ~,  
\end{equation}
which are the defining commutation relations of $K_2$ algebra with infinite 
growth, \cite{kac}; see also reference \cite{bakas3} for an alternative algebraic 
formulation, as well as reference \cite{finley} for earlier work on the subject 
using prolongation methods. 

Finally, it should be noted that although the zero curvature formulation of the 
problem requires the use of the additional element $H_0$, its integration by group 
theoretical methods employs only the $K_2$ subalgebra. Detailed analysis of the 
Toda system with $K=1$ and $S = \partial / \partial \theta$ shows that the 
general solutions assumes the final form 
\begin{equation}
\Phi (z, \bar{z}) = \Phi_0 (z, \bar{z}) - {\rm log} \left(<0| M_+^{-1}(z) 
M_-(\bar{z}) |0> \right) 
\end{equation}
where $|0>$ is a (formal) vacuum state of the $K_2$ algebra, 
\begin{equation}
X_i^+ |0> = 0 ~, ~~~~~ <0| X_i^- = 0 ~, ~~~~~ H_1 |0> = |0> ~, 
\end{equation}    
and $M_{\pm}$ are the path-ordered exponentials 
\begin{eqnarray}
M_+ (z) & = & {\cal P} {\rm exp} \left(\int^z dz^{\prime} e^{f(z^{\prime})} 
[X_1^+ + \psi (z^{\prime}) X_0^+] \right) , \nonumber\\
M_- (\bar{z}) & = & {\cal P} {\rm exp} 
\left(\int^{\bar{z}} d{\bar{z}}^{\prime} e^{\bar{f}({\bar{z}}^{\prime})} 
[X_1^- + \bar{\psi} ({\bar{z}}^{\prime}) X_0^-] \right)  
\end{eqnarray}
expressed in terms of the free fields $\Phi_0 (z, \bar{z}) = f(z) + \bar{f}(z, \bar{z})$ 
and $\Psi_0 (z, \bar{z}) = \psi(z) + \bar{\psi} (\bar{z})$. In geometrical terms, 
$\Psi_0 (z, \bar{z})$ prescribes the curvature of the two-dimensional metric in question, 
as $\Delta \Phi = \Psi_0 = -R$, whereas $\Phi_0$ will parametrize the general solution.   
In the simple case of axially symmetric metrics with $\Phi_0 = c \cdot (z + \bar{z}) +d$ 
and $\Psi_0 = a \cdot (z + \bar{z}) + b$, explicit calculation yields the power series 
expansion of the solution  
\begin{equation}
\Phi = \Phi_0 + {e^{\Phi_0} \over (1! ~c)^2} \left(\Psi_0 - {2a \over c}\right) 
+{e^{2\Phi_0} \over (2! ~c^2)^2} \left(\Psi_0^2 
-{4a \over c} \Psi_0 + {7a^2 \over 2c^2} \right) + {\cal O}(e^{3\Phi_0}) ~.   
\end{equation}      
Further details can also be found in the published work \cite{bakas4}. 

At this point we mention one more occurrence of the equation $\Delta \Delta \Phi = 0$ 
in the context of  
statistical mechanics. Recall first 
another fourth order parabolic  
equation of the form   
\begin{equation}
{\partial \over \partial t} e^{\Phi} = - \partial \bar{\partial} \left(e^{\Phi} 
\partial \bar{\partial} \Phi \right) , 
\end{equation}
which is a variant of the Calabi flow in that it involves the factor $e^{\Phi}$ rather 
than $e^{-\Phi}$ on its right-hand side; as such, it does not have a natural 
geometric interpretation. 
This equation, or better to say a one-dimensional reduction of it, 
arose while studying properties of interfaces between stationary 
phases of the two-dimensional (unbiased) Toom model, which are not described by 
equilibrium Gibbs ensembles, \cite{derrida}.     
It has the remarkable property to admit a fundamental solution 
of Gaussian type, 
\begin{equation}
e^{\Phi (z, \bar{z}; t)} = 
{1 \over 2\pi \sigma (t)} {\rm exp} \left(-{z \bar{z} \over 2 \sigma(t)} 
\right) ~,   
\end{equation}
with $\sigma (t) = \sqrt{t/2}$. It is similar to the fundamental solution 
of the second order heat equation,  
but differs from it in the $t$-dependence of $\sigma (t)$ that spreads as $\sqrt{t}$ 
and not $t$. The Calabi flow does not admit such solutions. However, 
static solutions of this equation are identical to static solutions of the Calabi 
flow, as the two are related by flipping the sign of $\Phi (z, \bar{z})$, and 
they are also characterized by the fourth order equation $\Delta \Delta \Phi = 0$.         
It will be interesting to examine further the physical manifestation of $K_2$ algebra 
in such models of statistical mechanics. 

In conclusion, we have reviewed the main aspects of 
geometric evolution equations as they arise in physics and mathematics with special 
emphasis on Ricci and Calabi flows in two dimensions. These are non-linear 
equations that are interrelated by methods of supersymmetric quantum mechanics 
and exhibit rich algebraic structure that enables us to cast them into zero 
curvature form. The infinite dimensional algebras that serve this purpose 
incorporate the deformation variable $t$ into their defining system of  
commutation relations and have infinite growth. It will be interesting in this 
context to consider further generalizations in low dimensions by including the effect of 
torsion in the corresponding geometric flows as well as non-perturbative effects 
generated by instantons, as in the quantum field theory of non-linear sigma 
models with Wess-Zumino-Witten terms and topological $\theta$-terms, respectively. 
Also, it will be interesting to examine the algebraic manifestation of 
entropy functionals for these flows in an attempt to study more systematically,  
and from a physics perspective, the new mathematical structures that were 
encountered above. 

Finally, other classes of geometric evolution equations, 
in particular those related to extrinsic curvature flows, deserve more 
attention in the future in view of their applications to quantum field 
theory and general relativity. Several applications to other branches of physics 
have been excluded from this presentation due to space limitations.      
There is also the hope that many more parabolic equations can be 
treated by methods similar to those developed here. Last, but not least, we 
mention the possible use of geometric evolution equations, such as the 
Ricci flow in $d$ spatial dimensions, for developing an off-shell formulation 
of the problem of gravitational collapse in $d+1$ space-time dimensions.   

\vskip1cm
\centerline{\bf Acknowledgments}

This work was supported in part by the European Research and Training Network 
``Constituents, Fundamental Forces and Symmetries of the Universe" under contract
number MRTN-CT-2004-005104 and the INTAS program ``Strings, Branes and Higher 
Spin Fields" under contract number 03-51-6346. I thank the Theory Division 
at CERN for hospitality and financial support during my sabbatical leave in the 
academic year 2004-05, where a big part of the present work was carried out in 
excellent and stimulating environment. I also thank the organizers of the AvH 
conference in Varna for their kind invitation to communicate these results to 
a wide audience.

\end{document}